\documentclass[prb,preprint]{revtex4-2} 
\usepackage[utf8]{inputenc}
\usepackage{amsmath}
\usepackage{comment}
\usepackage{graphicx}
\usepackage{endnotes}
\usepackage{amsfonts}
\usepackage{natbib}

\begin{document}




\title{On the Normalization and Density of 1D Scattering States}
\author{Chris L. Lin} 
\affiliation{Department of Physics, University of Houston, Houston, TX 77204-5005}

\date{\today}

\setlength\parindent{0pt}

\begin{abstract}

The normalization of scattering states is more than a rote step necessary to calculate expectation values. This normalization actually contains important information regarding the density of the scattering spectrum (along with useful details on the bound states). For many applications, this information is more useful than the wavefunctions themselves. In this paper we show that this correspondence between scattering state normalization and the density of states is a consequence of the completeness relation, and we present formulas for calculating the density of states which are applicable to certain potentials. We then apply these formulas to the delta function potential and the square well. We then illustrate how the density of states can be used to calculate the partition function for a system of two particles with a point-like (delta potential) interaction.
\end{abstract}

\maketitle

\section{Introduction}

In nonrelativistic quantum mechanics, eigenstates of the free Hamiltonian $\hat{H}_0 = \frac{\hat{P}^2}{2m}$ are also momentum eigenstates $|k \rangle$. These states satisfy the completeness relation

\begin{equation} \label{eqa3}
    \hat{I}=\int^\infty_{-\infty} \frac{dk}{2 \pi} \, |k \rangle   \langle k|,
\end{equation}

where $\hat{I}$ is the identity operator, and the normalization is $\langle q|k \rangle=2 \pi \delta(q-k)$ assuming the convention $\langle x| k \rangle \equiv \phi_k(x)=e^{ikx}$.
In the presence of a potential $\hat{V}$, the Hamiltonian $\hat{H}=\hat{H}_0+\hat{V}$ may possess bound states $|\psi_{bi} \rangle$ as well as scattering states $|\psi_k \rangle$, which are solutions to $\hat{H} |\psi_k \rangle=\frac{\hbar ^2 k^2}{2m} |\psi_k \rangle$. These scattering states are the sum of the free wave and a scattered part, i.e., $|\psi_k \rangle=|k \rangle+|\psi^{\text{scat}}_k \rangle$. The completeness relation then takes the form \cite{pat}:

\begin{align} \label{eq3}
    \hat{I}&=\sum_i |\psi_{bi} \rangle \langle \psi_{bi}|+\int^\infty_{-\infty} \frac{dk}{2 \pi} |\psi_k \rangle \langle \psi_k |.
\end{align}

Since the trace of the identity matrix gives the dimension of the underlying vector space, we can set the trace of Eq. \eqref{eqa3} equal to the trace of Eq. \eqref{eq3} giving:

\begin{align} \label{newEq1a}
    N_b&= -\int^\infty_{-\infty} dk \, \Delta \rho(k) \\ \nonumber
     \Delta \rho(k)&=\frac{1}{2\pi}\int^\infty_{-\infty} dx\, \left( |\psi_k(x)|^2-|\phi_k(x)|^2\right),
\end{align}

where $N_b$ is the number of bound states and we have defined $\Delta \rho(k)$ as the change in the momentum density of scattering states from that of the free states. Intuitively, Eq. \eqref{newEq1a} reflects the fact that that an increase in the number of bound states necessitates a decrease in the number of scattering states \cite{endnote1}. We will sometimes evaluate $\Delta \rho(k)$ by calculating a more general quantity

\begin{align} \label{newEq1}
    \Delta \rho(k,q)&=\frac{1}{2\pi}\int^\infty_{-\infty} dx\, \left( \psi^*_q(x)\psi_k(x)-\phi^*_q(x) \phi_k(x)\right),
     \end{align}
     
from which $\Delta \rho(k)$ can be obtained by setting $q=k$.\\

The fact that $\psi_k(x)$ and $\phi_k(x)$ have different normalizations may come as a surprise, although this has been observed in other papers \cite{stone2009mathematics, poliat, newton, Poliatzky:1994eh}. We will explicitly derive the difference in their normalizations in multiple ways and for multiple systems. Due to this correspondence with the density of states, the normalization of the scattering states is not merely an academic exercise. 

It is instructive to briefly outline how Eq. \eqref{newEq1a} can be obtained starting from a system with a countable discrete spectrum where both the bound and scattering states can be normalized to unity. This can be accomplished for instance by confining the interacting system to a box of length $L$ with periodic boundary conditions. The allowed energies and momenta can then be labeled by a quantum number $n$ and are therefore discrete satisfying $E(n,L) =  \frac{\hbar^2 k_n^2} {2m}$. For large $L$, increasing $n$ by 1 only causes an infinitesimal increment in $k$, so a sum over states can be turned into an integral by the sequence $\sum_n \,... \rightarrow \int dn \,... \rightarrow \int dk \, \frac{dn}{dk}\,...=\int dk \, \rho(k)\,...$. In particular, the completeness relation can be written \cite{footnoteA}:


\begin{align} \label{eq6z}
    \hat{I}&=\sum_i |\psi_{bi} \rangle \langle \psi_{bi}|+\sum_n |\psi_{k_n} \rangle \langle \psi_{k_n}|\\
&=\sum_i |\psi_{bi} \rangle \langle \psi_{bi}|+\int dk\, \rho(k) |\psi_{k_n} \rangle \langle \psi_{k_n}| \nonumber.
\end{align}

By defining the continuum spectrum solution as $|\psi_k \rangle=\sqrt{2\pi \rho(k)}|\psi_{k_n} \rangle$ we accomplish three things: Eq. \eqref{eq6z} goes into its continuum version Eq. \eqref{eq3}; $\langle \psi_k|\psi_k \rangle=2\pi\rho(k)\langle \psi_{k_n} |\psi_{k_n} \rangle =2\pi\rho(k)$ which gives Eq. \eqref{newEq1a}; $|\psi_k \rangle$ receives the correct units of square root of inverse momentum as those are the units of $\sqrt{2\pi \rho(k)}$ (by contrast, $|\psi_{k_n} \rangle$ is dimensionless). Although in principle one can calculate the density of states $\rho(k)$ by putting the system in a box, imposing boundary conditions, solving the Schr\"{o}dinger equation for $k=k(n,L)$ and then inverting ($n=n(k,L)$) to calculate $\rho(k)=\frac{dn}{dk}$, in this paper we will not discretize the system and instead directly use Eq. \eqref{newEq1a} to calculate $\Delta \rho(k)$.  

In section \ref{sec3}, we derive formulas for the normalization of the states in 1D, finite-range symmetric potentials. In section \ref{sec4} we apply these formulas to the delta function potential and the square well, and plot the density of states. In section \ref{sec6} we use the density of states to calculate the partition function for a system of two particles with a delta potential interaction.

\section{Derivation of the Change in the Density of States} \label{sec3}

We consider a symmetric potential $V(x)=V(-x)$ that is zero outside a region $|x|>a$. Symmetry allows us to consider only right-moving waves ($k>0$) since $\Delta \rho(k)=\Delta \rho(-k)$. The scattering state $|\psi_k \rangle$ is then given by

\begin{align}\label{waveEqnForm1}
\psi_k(x)=
\begin{cases}
e^{ikx}+R(k)\,e^{-ikx} & x<-a\\
\psi_{kI}(x) & -a \leq x \leq a\\
T(k)\, e^{ikx} & x>a,
\end{cases}
\end{align}

where $\psi_{kI}(x)$ is the wavefunction inside the range of the potential. First, we will place the system in a box of length $L$ and then consider the $L \rightarrow \infty$ limit \cite{endnote2}. Plugging Eq. \eqref{waveEqnForm1} into Eq. \eqref{newEq1} directly taking $q = k$ in the integrand, and using $|R(k)|^2+|T(k)|^2=1$, it is straightforward to show that

\begin{multline}
    \Delta \rho(k) =-\frac{a}{\pi}+\frac{1}{2\pi}\int_{-a}^{a} dx\,\psi^*_{kI}(x) \psi_{kI}(x)
    \\ -\frac{1}{2\pi k }\big(\text{Re}[R(k)]\sin(2ka)+\text{Im}[R(k)]\cos(2ka) \big)  \\
    +\frac{\sin(kL)}{2\pi k}\text{Re}[R(k)]+\frac{\cos(kL)}{2 \pi k}\text{Im}[R(k)]. 
\end{multline}

Both $\sin(kL)$ and $\cos(kL)$ oscillate very fast when $L \rightarrow \infty$, so that when $\Delta \rho(k)$ is integrated over $k$, these terms will produce zero unless their coefficients are non-analytic. The coefficients are analytic except at $k=0$. However, it is a general feature of scattering in 1D that $R(0)=-1$, i.e. complete inversion with no imaginary part \cite{doi:10.1119/1.15359,doi:10.1119/1.18123}, and reflects the fact that $k=0$ constitutes the opening of the continuum channel \cite{endnote3}. Using that $\frac{\sin(k L)}{k} \rightarrow\pi \delta(k)$ as $L\rightarrow \infty$ we get:
%
%
\begin{equation} \label{theDensityofDos}
    \Delta \rho(k)=-\frac{a}{\pi} -\frac{\text{Re}[R(k)]\sin(2ka)+\text{Im}[R(k)]\cos(2ka) }{2 \pi k}
    - \frac{1}{2} \delta(k) +\frac{1}{2\pi}\int_{-a}^{a} \,\left| \psi_{kI}(x) \right|^2 dx.
\end{equation}

We will comment on the $-\frac{1}{2} \delta(k)$ term in the next section; such a term is also found in studies of Levinson's theorem in 1D \cite{doi:10.1063/1.530481, Barton_1985}. \\

The integral in Eq. \eqref{theDensityofDos}  contains the wavefunction $\psi_{kI}(x)$ in the interaction region. An alternate formula for $\Delta \rho(k)$ that does not require knowledge of this wave function is:

\begin{equation}\label{shortcutEqn}
    \Delta \rho(k)=-\frac{i}{2\pi} \left(R^*(k)\frac{dR(k)}{dk}+T^*(k)\frac{dT(k)}{dk} \right)-\frac{1}{2} \delta(k).
\end{equation}

We briefly outline the derivation of Eq. \eqref{shortcutEqn} in appendix \ref{appendix1}, while in appendix \ref{appendix2} we will derive both Eq. \eqref{theDensityofDos} and Eq. \eqref{shortcutEqn} using a convergence factor in place of a box.

\section{Examples} \label{sec4}
\subsection{$\delta(x)$ Potential}

For the potential $V(x)=-g \delta(x)$ the reflection coefficient is $R(k) = -\left(1 + i k /\kappa \right)^{-1}$, where $\kappa=mg/\hbar^2$. This is valid for both the delta function well ($g>0$) and the delta function barrier $(g<0)$. When $g>0$, $\kappa$ is equal to $\sqrt{2 m E_b / \hbar^2}$ for the single bound state energy $E = -E_b = -m g^2 / 2\hbar^2$. By setting $a \rightarrow 0$ in Eq. \eqref{theDensityofDos} and plugging in $R(k)$, we get:

\begin{equation} \label{forDeltaDos}
    \Delta \rho(k)=-\frac{1}{2\pi}\frac{\kappa}{\kappa^2+k^2}-\frac{1}{2}\delta(k),
\end{equation}

\begin{figure}
    \centering
    \includegraphics[scale=.6]{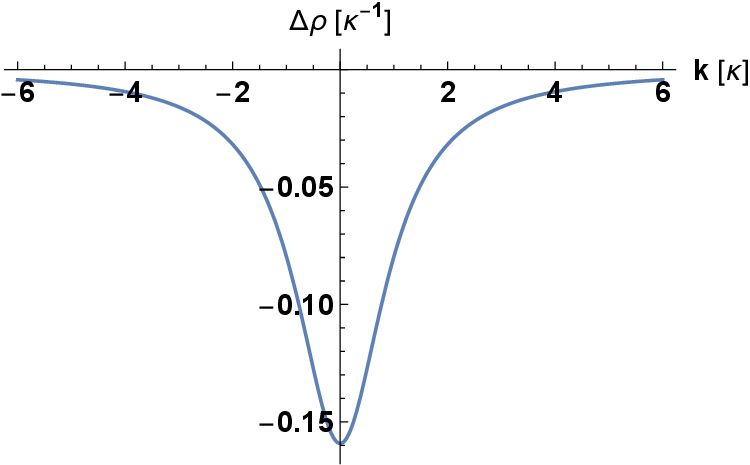}
    \caption{The change in the density of states for the potential $V(x)=-g \delta(x)$, assuming $g>0$. Not shown is a $-\frac{1}{2} \delta(k)$ term in $\Delta \rho(k)$ that only affects the $k=0$ state.}
    \label{figdosDelta}
\end{figure}

which is plotted in Fig. \ref{figdosDelta}. For the case of a barrier ($g < 0$), $\kappa<0$, so the graph in Fig. \ref{figdosDelta} would be inverted, implying all states except the $k=0$ state would experience an increase in density.\\

As $k \rightarrow \infty$,  $\Delta \rho(k) \rightarrow -\kappa / 2 \pi k^2$. It is interesting to note that $\kappa$, which is a measure of the bound state depth when $g > 0$, is the coefficient of the $k^{-2}$ decaying tail, which suggests a deeper bound state influences the spectrum at large $k$ more than a shallower bound state. 


Finally we verify the preservation of the size of the vector space:

\begin{align}
    N_b=-\int_{-\infty}^{\infty} dk\, \left(-\frac{1}{2\pi}\frac{\kappa}{\kappa^2+k^2}-\frac{1}{2}\delta(k) \right)=\begin{cases}
    1& \kappa>0\\
    0& \kappa<0.
    \end{cases}
\end{align}

\subsection{Square Well}

For a right-moving wave in a square well of length $2a$ and depth $V_0$ we express the results in terms of the transmission coefficient

\begin{align}
    T(k)&=\frac{e^{-2 i ka}}{\cos(2 \ell a)-i\frac{k^2+\ell^2}{2 k \ell}\sin(2 \ell a)},
    \end{align}
where $\ell=\sqrt{k^2+q^2}$, $q^2=2m V_0 / \hbar^2$ \cite{griffiths2018introduction}. The reflection coefficient and the wavefunction in the interacting region can then be expressed as
    \begin{align}
    R(k)&=\frac{i \sin(2 \ell a)(\ell^2-k^2)}{2 k \ell}\,T(k)  \\ \nonumber
    \psi_{kI}&=C \sin(\ell x)+D \cos(\ell x)  \\ \nonumber
    C&=[\sin(\ell a)+i \frac{k}{\ell}\cos(\ell a)]e^{ika}\,T(k) \\ \nonumber
    D&=[\cos(\ell a)-i \frac{k}{\ell}\sin(\ell a)]e^{ika}\,T(k).
\end{align}

\begin{figure}
    \centering
     \includegraphics[scale=0.7]{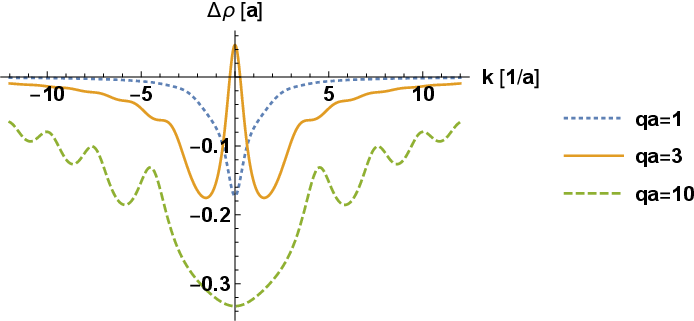}
     \caption{The change in the density of states for the square well, plotted for various values of $qa=\sqrt{\frac{2m V_0}{\hbar^2}}a$. Not shown is a $-\frac{1}{2} \delta(k)$ term in $\Delta \rho(k)$ that only affects the $k=0$ state.}
    \label{figdosSquare}
\end{figure}

Plugging these into Eq. \eqref{theDensityofDos} we get:

\begin{align}
\Delta \rho(k)=\frac{1}{2\pi}\left(-2a+ \frac{8 a k^2 \left(2 k^2+q^2\right)-\frac{2 q^4 \sin \left(4 a \sqrt{k^2+q^2}\right)}{\sqrt{k^2+q^2}}}{-q^4 \cos \left(4 a \sqrt{k^2+q^2}\right)+8 k^4+8 k^2 q^2+q^4} \right)-\frac{1}{2}\delta(k),
\end{align}

which is plotted in Fig. \ref{figdosSquare}. \\

A large $qa$ signifies a deep and wide well which allows for many bound states -- correspondingly, the area above the curves increases with $qa$. It is straightforward to show that asymptotically $\Delta \rho(k) \rightarrow -a q^2/(2 \pi k^2)$ as $k \rightarrow \infty$. Once again the coefficient of the $k^{-2}$ decaying tail is related to the depth of the well.\\

Finally, using the substitution $u = k a$, we calculate the number of bound states by integrating the density of states:

\begin{align} \label{intermediateBoundeqn}
    N_b= \frac{1}{\pi} \int^\infty_{0} du\,\left(2 - \frac{8  u^2 \left(2 u^2+(qa)^2\right)-\frac{2 (qa)^4 \sin \left(4 \sqrt{u^2+(qa)^2}\right)}{\sqrt{u^2+(qa)^2}}}{8 u^4+8 u^2 (qa)^2+(qa)^4 - (qa)^4 \cos \left(4  \sqrt{u^2+(qa)^2}\right)} \right)+\frac{1}{2}.
\end{align}

\begin{figure}
    \centering
    \includegraphics[scale=.7]{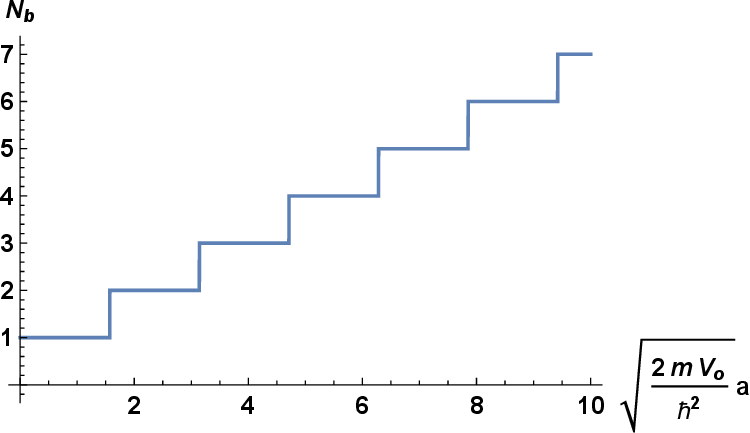}
    \caption{The number of bound states in the square well, found by numerical integration of the change of the density of states, $\Delta \rho(k)$. The well always has one bound state no matter how small the attraction, and gains an additional bound state whenever $\sqrt{\frac{2m V_0}{\hbar^2}}a$ increases by $\frac{\pi}{2}$.}
    \label{boundPlot}
\end{figure}

The result is plotted in Fig. \ref{boundPlot} for various values of $qa$. An analysis of the bound state sector \cite{williams2003topics} gives the number of bound states as $N_b=1+\lfloor \frac{2}{\pi} (qa) \rfloor$, where the floor function $\lfloor \frac{2}{\pi} (qa) \rfloor$ returns the greatest integer that is smaller than $\frac{2}{\pi} (qa)$. It is remarkable that a numerical integration of Eq. \eqref{intermediateBoundeqn} can produce the sharp steps seen in Fig. \ref{boundPlot}. 

\section{Application to a Thermodynamic System}\label{sec6}

In this section we illustrate how this technique of calculating the density of states can be applied to a physical system in thermal equilibrium. Consider a system of two particles of mass $m$ confined to a line of length $L$ and in thermal contact with a heat bath at temperature $T$.  The particles interact with each other through the contact interaction $V(x_1, x_2) = -g \delta(x_1 - x_2)$.   The Hamiltonian for such a system is $H=\frac{p_1^2}{2m}+\frac{p_2^2}{2m}-g \delta(x_1-x_2)$. Changing to center of mass and relative coordinates the Hamiltonian becomes separable ($H=H_{\text{com}}+H_{\text{rel}}$) with

\begin{align}
    H_{\text{com}}&=\frac{P^2}{4m}\\ \nonumber
    H_{\text{rel}}&=\frac{p^2}{m}-g \delta(x),
\end{align}

where $P=p_1+p_2$, $p=\frac{1}{2}\left(p_1-p_2\right)$, and $x=x_1-x_2$ are canonical variables. The partition function for this system can then be factorized as $Q_2=Q_2^{\text{rel}}Q_2^{\text{com}}$\cite{endnote4}. $H_{\text{com}}$ has the form of a Hamiltonian for a free particle of mass $2m$ so its partition function is

\begin{align}
    Q^{\text{com}}_2=\int^\infty_{-\infty}dk\, \rho_\text{free} \, e^{-\beta \frac{\hbar^2 k^2}{4m}}=L \sqrt{\frac{m}{\pi \beta \hbar^2}},
\end{align}

where we inserted $\rho_\text{free}=\frac{L}{2\pi}$, the density of states for the free particle \cite{endnote5}.  The relative two-particle partition function can be calculated as 

\begin{align} \label{eqnPart}
    Q^{\text{rel}}_2=\sum_i e^{-\beta E_{bi}}+\int^\infty_{-\infty} dk \left(\rho_{\text{free}}+\Delta \rho(k) \right)e^{-\beta \frac{\hbar^2 k^2}{2 \mu}},
\end{align}
where $\mu$ is the reduced mass, $\beta=(k_B T)^{-1}$ (where $k_B$ is Boltzmann's constant), and $E_{bi}$ are bound state energies (if they exist for the system).  We calculate $Q_2^{\text{rel}}$ by integrating the scattering spectrum over the density of states. Inserting Eq. \eqref{forDeltaDos} into Eq. \eqref{eqnPart} yields the two-particle relative partition function:

\begin{align}\label{zzequat19}
    Q_2^{\text{rel}}&=\left[e^{-\beta \left(-\frac{\hbar ^2\kappa^2}{m}\right)}+\int^\infty_{-\infty} dk \left(\frac{L}{2\pi}-\frac{1}{2\pi}\frac{\kappa}{\kappa^2+k^2}-\frac{1}{2}\delta(k)\right)e^{-\beta \frac{\hbar^2 k^2}{m}}\right] \\ \nonumber
    &
    =\frac{1}{2}\left\{L \sqrt{\frac{m}{\pi \beta \hbar^2}}+
	e^{\frac{\beta \hbar^2 \kappa^2}{m}}
    \left[
    \text{erf}\left(\hbar \kappa \sqrt{\frac{\beta}{m}}\right)+1
    \right]
    - 1
    \right\}
    ,
\end{align}

where $\text{erf}(x)$ is the error function. If the interaction is very weak ($\kappa \rightarrow 0$), $Q_2^{\text{rel}}$ approaches the partition function of a free particle of mass $m/2$: the bound state term would exactly cancel the scattering term in Eq. \eqref{zzequat19}. This provides a nontrivial check of Eq. \eqref{theDensityofDos} which was used to derive the scattering term.\\

\section{Conclusions}

We have shown that the density of scattering states lies hidden in the normalization of the scattering wave functions. Moreover, the scattering state sector contains information about the bound state sector. The connection between bound states and scattering states is usually presented to students through either Levinson's theorem, \cite{levl} or through a discussion of the analytic properties of the scattering matrix \cite{Sakurai:1167961}. In this paper we highlighted this connection using an alternative approach which also makes contact with the density of states. Because the density of states is an important concept in statistical mechanics, we find this approach to be pedagogically beneficial and have used it to calculate the exact partition function of an interacting system.

\section{Conflict of Interests}

The author has no conflicts to disclose.

\section*{Acknowledgements}

The author would like to acknowledge the reviewers for their helpful suggestions.


\begin{thebibliography}{20}
\bibitem [{end({\natexlab{a}})}]{endnote1}%
  \BibitemOpen
  \href@noop {} { {\bibinfo {title} {In linear algebra, two different
  basis $\{|k \rangle \}$ and $\{|\phi_{bi}\rangle, \, |\psi_k \rangle \}$
  should have the same size, so the more states $\{|\phi_{bi}\rangle\}$ that
  exist the less the density of the $\{|\psi_k \rangle\}$ states compared to
  the density of the $\{|k \rangle \}$ states}}\ } \BibitemShut
  {NoStop}%
\bibitem{footnoteA} An interval of size $\Delta k$ centered around the momentum $k$, may contain
several discrete momenta $\{..., k_{n-1}, k_n, k_{n+1},... \}$. Instead of adding them
$...+|\psi_{k_{n-1}} \rangle \langle \psi_{k_{n-1}}|+
|\psi_{k_{n}} \rangle \langle \psi_{k_{n}}|+|\psi_{k_{n+1}}
\rangle \langle \psi_{k_{n+1}}|+...$,
we take any one of the discrete momenta as representative
and multiply by the number of elements in the set:\\
\begin{equation*}
... + \vert \psi_{k_{n-1}} \rangle \langle \psi_{k_{n-1}}|+|\psi_{k_{n}} \rangle \langle
\psi_{k_{n}}|+|\psi_{k_{n+1}} \rangle \langle \psi_{k_{n+1}}|+...
=\rho(k) \Delta k |\psi_{k_{n}} \rangle \langle \psi_{k_{n}}|
\end{equation*}\\
Therefore the $|\psi_{k_{n}} \rangle$ in the last term of Eq. (\ref{eq6z}) represents
any of the discrete states that fall within the interval $dk$ centered
around the momentum $k$.
\bibitem [{end({\natexlab{b}})}]{endnote2}%
  \BibitemOpen
  \href@noop {} { {\bibinfo {title} {We will not impose any boundary
  conditions at the endpoints of the box as we are not attempting to discretize
  the spectrum, but rather regulate the infinite volume limit}}\ }
  \BibitemShut {NoStop}%
\bibitem [{end({\natexlab{c}})}]{endnote3}%
  \BibitemOpen
  \href@noop {} { {\bibinfo {title} {For exotic cases where there
  exists a zero-energy bound state, we may have $R(0) \neq -1$, which is known
  as threshold anomaly. We do not consider such cases here}}\ }
  \BibitemShut {NoStop}%
\bibitem [{end({\natexlab{d}})}]{endnote4}%
  \BibitemOpen
  \href@noop {} { {\bibinfo {title} {When the energy of a system of two
  particles decomposes independently into $E^{\text{com}}_i+E^{\text{rel}}_i$,
  then the partition function factorizes: $Q_2=\sum_{I,i} e^{-\beta
  \left(E^{\text{com}}_I+E^{\text{rel}}_i\right)}=\left(\sum_{I} e^{-\beta
  E^{\text{com}}_I}\right) \left(\sum_{i} e^{-\beta
  E^{\text{rel}}_i}\right)=Q_2^{\text{com}} Q_2^{\text{rel}}$}}\ }
  \BibitemShut {NoStop}%
\bibitem [{end({\natexlab{e}})}]{endnote5}%
  \BibitemOpen
  \href@noop {} {{\bibinfo {title} {If we impose periodic boundary
  conditions on a free particle in a box of length $L$, we get the condition
  $k(n)=\frac{2\pi n}{L}$. The density is then
  $\rho(k)=\frac{dn}{dk}=\frac{L}{2\pi}$}}\ } \BibitemShut
  {NoStop}%
\end{thebibliography}


\appendix
\section{Asymptotic Form} \label{appendix1}

The derivation of Eq. \eqref{shortcutEqn} is challenging. Consider the Schr\"{o}dinger equation and its conjugate:

\begin{align}
    -\frac{\hbar^2}{2m}\frac{d^2}{dx^2} \psi_k+V\psi_k&=\frac{\hbar^2 k^2}{2m} \psi_k\\ \nonumber
     -\frac{\hbar^2}{2m}\frac{d^2}{dx^2} \psi^*_q+V\psi^*_q&=\frac{\hbar^2 q^2}{2m} \psi^*_q.
\end{align}

Multiplying the top line by $\psi^*_q$ and the bottom line by $\psi_k$ and subtracting removes reference to the potential:

\begin{align}
    \psi^*_q \psi_k=\frac{1}{q^2-k^2}\left(\psi^*_q \frac{d^2 \psi_k}{dx^2}-\psi_k \frac{d^2 \psi^*_q}{dx^2} \right).
\end{align}

We can use integration by parts to get an expression involving the wavefunction only in the asymptotic region:

\begin{align} \label{equat19}
    \int^{L/2}_{-L/2} dx\, \psi^*_q \psi_k=\frac{1}{q^2-k^2}\left(\psi^*_q \frac{d \psi_k}{dx}-\psi_k \frac{d \psi^*_q}{dx} \right)\Bigg |^{L/2}_{-L/2}\equiv \frac{N(k,q)}{q^2-k^2}.
\end{align}

Since $L$ is large, the asymptotic forms $\psi_k(L/2)=T(k)e^{i k (L/2)}$, $\psi'_k(L/2)=ik T(k)e^{i k (L/2)}$, $\psi_k(-L/2)=e^{i k (-L/2)}+R(k) e^{-ik (-L/2)}$, etc. can be used to find the numerator $N(k,q)$. Plugging Eq. \eqref{equat19} into Eq. \eqref{newEq1} and taking $q \rightarrow k$ gives $\Delta \rho(k)$. L'H\^{o}pital's rule can be used to evaluate the limit as


\begin{align} \label{equat19a1}
  \lim_{q \rightarrow k} \frac{N(k,q)}{q^2-k^2}= \frac{ \partial_q N(k,q) |_{q=k}}{2k},
\end{align}
which creates the derivatives seen in Eq. \eqref{shortcutEqn}. After the limit is taken, the conservation of probability and its derivative 

\begin{align} 
  R^*(k)R(k)+T^*(k)T(k)&=1 \\ \nonumber
  \frac{dR^*}{dk}R+\frac{dT^*}{dk}T&=-\left(\frac{dR}{dk}R^*+\frac{dT}{dk}T^* \right)
\end{align}

can be used to further simplify the expression to get the final form in Eq. \eqref{shortcutEqn}.

\section{Convergence Factor Derivation} \label{appendix2}

Without a box, we will utilize the help of a convergence factor ($\epsilon>0$) that converts oscillatory expressions into decaying ones:

\begin{align} \label{aaaab}
    \Delta \rho(k,q)&=\frac{1}{2\pi}\lim_{\epsilon \rightarrow 0^+} \int_{-\infty}^\infty dx\, \left( \psi^*_q(x)\psi_k(x)-\phi^*_q(x) \phi_k(x)\right)e^{-\epsilon |x|}.
\end{align}

Inserting Eq. \eqref{waveEqnForm1} into Eq. \eqref{aaaab} all integrals in the asymptotic region $|x|>a$ are of the form $\int^\infty_a dx \, e^{isx}e^{-\epsilon x}=\frac{-e^{isa}e^{-\epsilon a}}{is-\epsilon}$ and $\int^{-a}_{-\infty} dx\, e^{isx}e^{\epsilon x}=\frac{e^{-isa}e^{-\epsilon a}}{is+\epsilon}$. Performing these integrals we get:

\begin{multline} \label{secondaryAppendix}
    \int_{-\infty}^\infty dx \, \psi^*_q(x)\psi_k(x)e^{-\epsilon |x|} =\left(T^*(q)T(k)+R^*(q)R(k) \right)\left(\frac{1}{i(k-q)+\epsilon}-\frac{1}{i(k-q)-\epsilon} \right) e^{i(k-q)a}e^{-\epsilon a}
    \\+ 
    \left(1-\left(T^*(q)T(k)+R^*(q)R(k)  \right)e^{2ia(k-q)} \right)\frac{e^{-ia(k-q)}e^{-\epsilon a}}{i(k-q)+\epsilon}\\+
    \left(R(k)\frac{e^{ia(k+q)}}{-i(k+q)+\epsilon}+R^*(q)\frac{e^{-ia(k+q)}}{i(k+q)+\epsilon}\right)e^{-\epsilon a}\\+
    \int_{-a}^a dx\, \psi^*_{qI}(x) \psi_{kI}(x) e^{-\epsilon |x|}.
\end{multline}

Setting $q=k$, the first term on the right-hand side produces $2\pi \delta(0)-2a$, the second term is $0$ from $|R(k)|^2+|T(k)|^2=1$, and the third term produces $-\frac{1}{k }\big(\text{Re}[R(k)]\sin(2ka)+\text{Im}[R(k)]\cos(2ka) \big)
    - \pi \delta(k)$. This is in agreement with Eq. \eqref{theDensityofDos} once the free normalization is subtracted, which gets rid of the $2\pi \delta(0)$ term.  \\
    
    On the other hand, if in Eq. \eqref{secondaryAppendix} we take $k$ close to, but not equal to $q$, then we can set $\epsilon=0$ to get a useful relation:
    \begin{multline} \label{lastEqn}
    0=0
    + \lim_{k \rightarrow q} 
    \left(1-\left(T^*(q)T(k)+R^*(q)R(k)  \right)e^{2ia(k-q)} \right)\frac{e^{-ia(k-q)}}{i(k-q)}\\
    -\frac{1}{q }\big(\text{Re}[R(q)]\sin(2qa)+\text{Im}[R(q)]\cos(2qa) \big) \\+
    \int_{-a}^a dx\, \psi^*_{qI}(x) \psi_{qI}(x).
\end{multline}

Using L'H\^{o}pital's rule on the first term on the right-hand of Eq. \eqref{lastEqn} gives $-2a+i T^*(q) \frac{dT(q)}{dq}+i R^*(q) \frac{dR(q)}{dq}$. This allows us to solve for $ \int_{-a}^a dx\, \psi^*_{qI}(x) \psi_{qI}(x)$ in terms of asymptotic quantities:

  \begin{multline} 
    \int_{-a}^a dx\, \psi^*_{qI}(x) \psi_{qI}(x)=2a-i T^*(q) \frac{dT(q)}{dq}-i R^*(q) \frac{dR(q)}{dq}
 \\
    +\frac{1}{q }\big(\text{Re}[R(q)]\sin(2qa)+\text{Im}[R(q)]\cos(2qa) \big),
\end{multline}

which we then subsequently plug into Eq. \eqref{theDensityofDos} to get Eq. \eqref{shortcutEqn}.

\end{document}